\documentclass{aa}
\usepackage{graphicx}
\begin{document}

\title{Equation of state of dense matter and the minimum mass of 
cold neutron stars}
\author{P. Haensel\inst{1} 
\and J.L. Zdunik\inst{1}
 \and
F. Douchin \inst{2,3}
}
\institute{N. Copernicus Astronomical Center, Polish
           Academy of Sciences, Bartycka 18, PL-00-716 Warszawa, Poland
\and
Department of Physics, University of Illinois at Urbana-Champaign, Urbana, 
Illinois 61801, U.S.A.
\and
Centre de Recherche Astronomique de Lyon, ENS de Lyon, 
69364 Lyon, France\\ 
\\
{~~~~~\tt haensel@camk.edu.pl, jlz@camk.edu.pl 
  }}
\offprints{P. Haensel}
\date{24 January, 2002
}
\abstract{
Equilibrium configurations of cold neutron stars 
near the minimum mass 
are studied, using the recent equation of state SLy, which describes 
in a unified, physically consistent manner, both the solid crust and 
the liquid core of neutron stars. Results are compared with those 
obtained using an older FPS equation of state of cold catalyzed matter. 
The value of $M_{\rm min}\simeq 0.09~{\rm M}_\odot$ depends very 
weakly on the equation of state of cold catalyzed matter: it is 
$0.094~{\rm M}_\odot$ for the SLy  model, and $0.088~{\rm M}_\odot$ 
for the FPS one. Central density at $M_{\rm min}$ is 
significantly lower than 
the normal nuclear density: for the SLy 
equation of state we get central density 
$1.7~10^{14}~{\rm g~cm^{-3}}$, to be 
compared with $2.3~10^{14}~{\rm g~cm^{-3}}$ obtained for the FPS 
one. Even at $M_{\rm min}$, neutron stars have a small liquid 
core of radius of about 4 km, 
containing some   2-3\% of  the stellar mass. Neutron stars 
with  
 $0.09~{\rm M}_\odot<M<0.17~{\rm M}_\odot$
 are bound with respect to dispersed 
configuration of the hydrogen gas, but are unbound with respect to 
dispersed $^{56}{\rm Fe}$. 
The effect of uniform rotation on the minimum-mass configuration of cold 
neutron stars is studied. Rotation increases the value of $M_{\rm min}$; 
at rotation period of 10 ms  the minimum mass of neutron stars 
increases to $0.13~{\rm M_\odot}$, and 
corresponds to the mass-shedding 
(Keplerian) configuration. In the case of the shortest observed 
rotation period of radio pulsars  1.56 ms, minimum mass of 
uniformly rotating cold neutron stars corresponds to the mass-shedding 
limit, and is found at
 $0.61~{\rm M}_\odot$ for the SLy EOS and $0.54~{\rm M}_\odot$ 
for the FPS EOS. 
}

\titlerunning{ Minimum mass of cold neutron stars}
\authorrunning{P. Haensel et al.}
\maketitle

\section{Introduction}
%
Mass of neutron stars is bound on the lower-density side by 
the minimum mass, $M_{\rm min}$. Equilibrium 
neutron star configurations 
with $M<M_{\rm min}$ do not exist. Some astrophysical scenarios, 
in which a neutron star looses matter, and reaches the critical 
value $M_{\rm min}$, were proposed (Blinnikov et al. 1984, 
Colpi et al. 1989, Colpi et al. 1991,  Sumiyoshi et al. 1998).  
 Basically, one considers a neutron star 
in a close binary system with a more compact, and more massive 
object (more massive neutron star, stellar mass black hole). At 
a sufficiently small separation between a neutron star and its 
companion,  
less massive, and therefore larger, neutron star starts 
to lose mass  due to gravitational pull of its companion, 
the lost matter being accreted by the companion compact object. 
This process is self-accelerating, because decrease of mass leads 
to increase of neutron star size, making it even more susceptible 
to the mass loss. After crossing $M_{\rm min}$, no equilibrium 
configuration can be reached. Numerical simulations suggest, 
that after crossing the $M_{\rm min}$ value, a neutron star will
undergo an explosion (Blinnikov  et al. 1984, Colpi et al. 1989, 1991, 
Sumiyoshi et al. 1998). 

Of course, rotation and thermal effects can affect  
(actually - increase) the actual value of the minimum neutron star mass. 
As we will see, rotation will also change the nature of the 
minimum-mass  configuration. 
In the present paper, we will restrict ourselves 
 to neutron stars built 
of cold catalyzed matter. For newly born  proto-neutron 
stars, both thermal and neutrino-trapping effects are 
large, and are expected to increse minimum mass to 
$0.9-1.1~{\rm M}_\odot$ (Goussard et al. 1998, 
Strobel et al. 1999).  
However, $M_{\rm min}$ for static, cold neutron stars  
still remains the absolute lower bound on the 
mass of neutron star. 

The calculation of $M_{\rm min}$ for neutron stars has a long 
history. The first correct estimate of $M_{\rm min}$ was obtained 
by Oppenheimer \& Serber (1938). They showed, that the 
previous estimate obtained by Landau (1938) was much 
too small, and moreover based on incorrect arguments. Oppenheimer 
\& Serber (1938) obtained an estimate $0.17~{\rm M}_\odot$ 
without taking into account nuclear interactions. Then they 
argued, that inclusion of the contribution from the nucleon-nucleon 
interactions leads to a decrease of $M_{\rm min}$ to $0.03-0.10~
{\rm M}_\odot$, depending on the assumptions they made about 
the (very poorly known  at that time) nucleon-nucleon 
interaction. It should be stressed, that their criterion of 
stability was based  on energy arguments: a neutron star had to 
be stable with respect to transformation into medium mass 
nuclei (they took calcium as an example). 
Calculation of Harrison et al. (1964) was based on the 
Harrison-Wheeler equation of state (Harrison et al. 1964), which 
described both the crust and the liquid core of neutron stars. 
From the present-day point of view, their equation of state above neutron 
drip was unrealistic. They obtained $M_{\rm min}=0.18~{\rm M}_\odot$, 
at a central density of only $3~10^{13}~{\rm g~cm^{-3}}$. 
Using two versions of the Levinger \& Simmons (1961) 
baryon-baryon potential, Tsuruta \& Cameron (1966) 
obtained $M_{\rm min}=0.11-0.13~{\rm M}_\odot$. 
Cohen \& Cameron (1971), using the equation of state (EOS) 
based on the up-dated version of the 
Levinger-Simmons nucleon-nucleon potential (Langer 
et al. 1969), obtained $M_{\rm min}=0.065~{\rm M}_\odot$. 
Finally, in their classical paper on the EOS and structure 
of neutron stars, Baym et al. (1971b, hereafter BPS), 
who used their own (BPS) EOS of the outer crust, 
and that of Baym et al. (1971a, hereafter BBP) for the 
inner crust and for the outer layers of the liquid core, 
obtained $M_{\rm min}=0.0925~{\rm M}_\odot$.

The properties of the equilibrium configurations  near 
$M_{\rm min}$ are sensitive to the EOS of neutron star matter 
at subnuclear densities, i.e. for mass density significantly 
below normal (saturation) density of nuclear matter, 
$\rho_0=2.7~10^{14}~{\rm g~cm^{-3}}$ (baryon density 
significantly below $n_0=0.16~{\rm fm^{-3}}$). 
As we will show, a particularly important role is played by 
the EOS near the inner crust - liquid core boundary. Therefore, 
the EOS should give physically correct description of this 
transition region; a brutal {\it ad hoc} matching of the inner 
crust and liquid core are not sound for this purpose. 
The microscopic model, underlying the EOS, should be 
the same on both sides of the crust-core boundary, and 
should therefore be based on the same effective nuclear hamiltonian. 

In the present paper, we calculate $M_{\rm min}$ and study 
static configurations near minimum mass. We also study  
the effect of rotation on these configurations. 
We use two EOS, 
based on different effective nuclear hamiltonians. 
These effective hamiltonians are:  a recent 
SLy model 
(Chabanat et al. 1998, for applications see 
Douchin \& Haensel 2000, Douchin et al. 2000, 
EOS in   Douchin \& Haensel 2001), and a somewhat 
older FPS model (Ravenhall \& Pandharipande 1989, 
for applications see Lorenz et al. 1993). 

The plan of the paper is as follows. Equation of state near
 crust-liquid core interface is discussed in Sect.2. In 
Sect. 3, we study equilibrium configurations near $M_{\rm min}$, 
and their dependence on the assumed EOS model. The problem 
of binding energy and stability of low-mass neutron stars 
is studied in some detail in Sect.4. 
Possible effects of elastic strain within the 
equilibrium configurations near $M_{\rm min}$ are 
estimated in Sect. 5. 
Effects of 
rotation are studied in Sect. 6. Finally, Sect. 7 
contains discussion of our results and  conclusions. 
%
\section{Equation of state of cold catalyzed matter near the crust-core 
interface
}
%
\begin{figure}
\resizebox{\hsize}{!}{\includegraphics{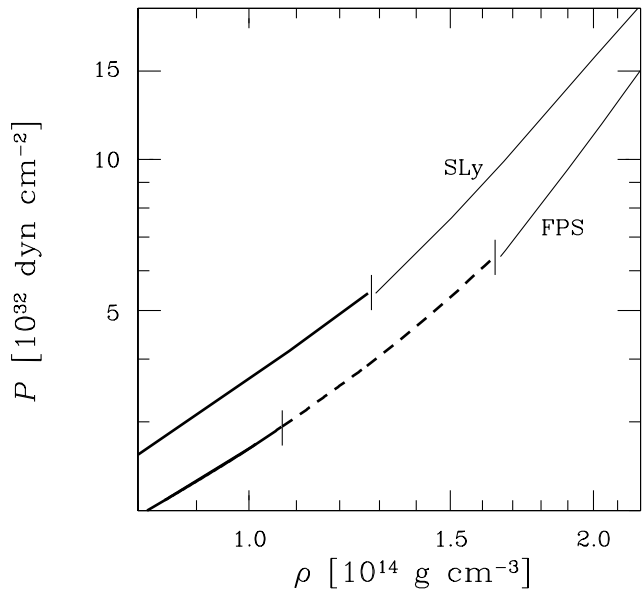}}
\caption{
Comparison of the SLy and FPS EOS near the crust-liquid core 
transition. Thick solid line: inner crust with spherical nuclei. 
Dashed line corresponds to non-spherical (``exotic'') nuclear 
shapes. Thin solid line: uniform $npe$ matter.  Vertical dashes 
separate different dense  matter phases.  
  }
 \label{EOS.cc}
\end{figure}
The properties of  neutron stars close to $M_{\rm min}$ are 
determined by the EOS of dense matter at subnuclear densities, 
and  in particular in the density interval 
$0.1\rho_0<\rho<\rho_0$. 
At such densities, the actual structure of the ground state of 
matter results from the interplay of the surface and Coulomb terms 
in the total energy density. These tiny ``finite size'' terms,  which 
are absent in uniform $npe$ matter,  
are difficult to calculate with high precision at the density 
of interest. However, these  terms, very small 
compared to the main bulk energy contribution, 
decide what is the 
actual shape of nuclei, and whether or not 
matter with nuclear structures 
is energetically preferred over the spatially uniform 
$npe$ matter. 

As we already stressed in Sect.1, it is mandatory 
 to use the same effective nuclear 
hamiltonian to describe both nuclear structures (nuclei) and 
neutron gas in the 
inner crust,  and the  uniform $npe$ matter of the liquid core. In the 
present paper we use a recently developed SLy model of neutron star 
matter (Douchin \& Haensel 2000, Douchin et al. 2000,  
Douchin \& Haensel 2001), based on the 
SLy effective nuclear interaction (Chabanat et al. 1997, 1998). 
 For this  
 model,  
 nuclei in the ground state of neutron star  matter 
remain spherical down to 
the  bottom edge of the inner crust (Douchin \& Haensel 2000). 
Transition to the uniform $npe$ plasma takes place  
at $\rho_{\rm edge}=1.3~10^{14}~{\rm g~cm^{-3}}$ (baryon density 
$n_{\rm edge}=0.078~{\rm fm^{-3}}$). The crust-core phase transition is 
very weakly first-order, with a relative density jump of about   1\%. 

Unfortunately, absence or presence of non-spherical 
(called also ``exotic'' or ``unusual'') 
nuclear shapes in the bottom layers of the inner crust  depends on the 
assumed model of effective nuclear hamiltonian. As an  
example of a different effective nuclear hamiltonian, we will 
consider the FPS model (Pandharipande \& Ravenhall 1989, 
Lorenz et al. 1993), which also gives a unified  description of both 
the inner crust and the liquid core. For the FPS model, the  
crust-liquid core transition takes place  at higher density 
$\rho_{\rm edge}=1.6~10^{14}~{\rm g~cm^{-3}}$ (baryon density 
$n_{\rm edge}=0.096~{\rm fm^{-3}}$), and is preceded by a sequence 
of phase transitions between various nuclear shapes (Lorenz et al. 1993). 
The sequence of phase transitions starts at 
$1.1~10^{14}~{\rm g~cm^{-3}}\simeq {1\over 3}\rho_0$ 
(baryon  density $n_{\rm b}=0.064~{\rm fm^{-3}}$), 
at which spherical nuclei immersed in neutron gas 
are replaced by the cylindrical ones. 
Then follows a transition from cylindrical nuclei 
immersed in neutron gas to the 
slabs of nuclear matter interspaced with neutron gas, 
which at still higher density are 
 replaced by nuclear matter with cylindrical holes 
filled with neutron gas,  replaced eventually by nuclear matter with 
spherical bubbles filled with neutron gas. This last 
phase persists up to $\rho_{\rm edge}$. 
All phase transitions are very weakly first-order, with relative 
density jumps smaller than 1\%. 

Both EOS are compared, in the relevant density interval, 
 in Fig.\ref{EOS.cc}. In the vicinity of the crust-core 
interface, the SLy EOS is stiffer than the FPS one. 
Moreover, in the case of the SLy EOS the discontinuous 
stiffening (jump of compression modulus) at the crust-core 
transition is more pronounced than in the FPS case, where 
it has been preceded by a sequence of phase transitions connected 
with changes of nuclear shapes. Such a sequence of 
phase transitions implied therefore 
a gradual stiffening of matter. 

\section{Mass, radius and  central density 
 of configuration close to $M_{\rm min}$}
%
\begin{figure}
\resizebox{\hsize}{!}{\includegraphics{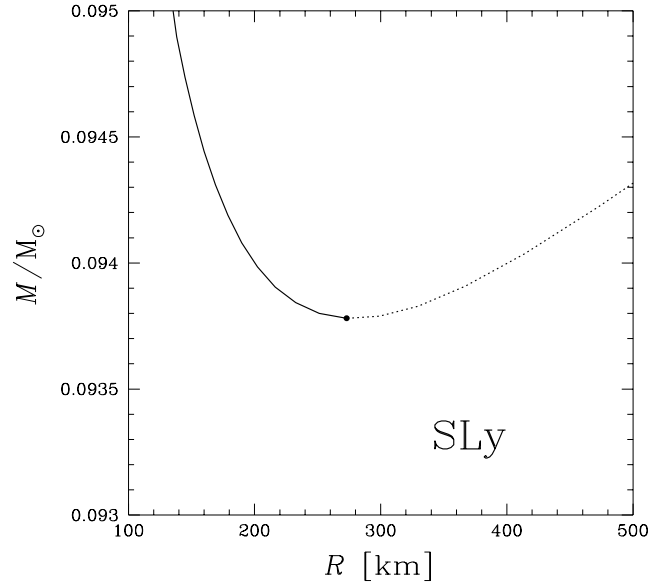}}
\caption{
Gravitational mass versus radius near  $M_{\rm min}$, for 
cold, static neutron stars. 
Calculation performed for  the SLy EOS. 
 Solid line - stable configurations, dotted lines - 
configurations unstable with respect to small radial perturbations. 
Minimum mass configuration is indicated with a filled circle.  
  }
 \label{MRmin}
\end{figure}
\begin{figure}
\resizebox{\hsize}{!}{\includegraphics{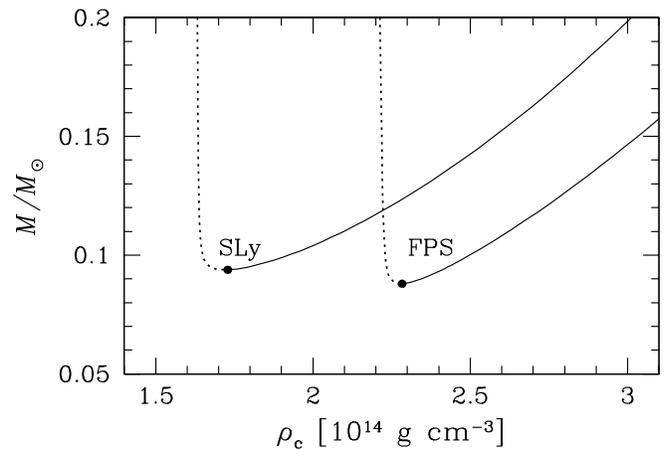}}
\caption{
Gravitational mass versus central density, in the vicinity 
of $M_{\rm min}$, for cold, static neutron stars. 
 Calculations performed for the SLy 
and the FPS EOS. Solid line - stable configurations, dotted lines - 
configurations unstable with respect to small radial perturbations. 
Minimum mass configuration is indicated with a filled circle.  
  }
 \label{Mrhoc}
\end{figure}
%
\begin{table*}
\caption{\label{Mmin.tab}
Configurations of minimum mass for cold, static neutron stars}
\begin{tabular}{|cccccccc|}
\hline
 EOS & $M_{\rm min}$ & $\rho_{\rm c}$ & $R$ &
 $M_{\rm core}/M$ & $R_{\rm core}$ & $E_{\rm bind}^{(\rm Fe)}$ 
& $E_{\rm bind}^{\rm (H)}$ \cr
  & $[{\rm M}_\odot]$ & $[10^{14}~{\rm g/cm^3}]$ & [km] &
 &    [km]  
& [$10^{51}$~erg]
& [$10^{51}$~erg]\cr
\hline
&&&&&&&\\
SLy  & 0.094  & 1.6  &  270  & 0.02 & 3.8 & -1.0  &  0.5\\
&&&&&&&\\
FPS  & 0.088  & 2.2  & 220 & 0.03 & 4.2  &  -1.2 & 0.2  \\
&&&&&&&\\
\hline
\end{tabular}
\end{table*}

The mass-radius plot  for the SLy EOS, in the vicinity of 
$M_{\rm min}$, is shown in Fig.\ref{MRmin}. The mass minimum 
is very flat, particularly on the lower-density branch. 
In contrast, the minimum of the mass-central density curve 
is very sharp, and the curve  $M(\rho_{\rm c})$ on the 
lower-density side is extremely steep, as seen in Fig.\ref{Mrhoc}. 
Configurations to the left of the minimum in Fig.\ref{Mrhoc} 
have ${\rm d}M/{\rm d}\rho_{\rm c}<0$, and are therefore 
unstable with respect 
to small radial perturbations; they are  not expected to exist 
in the Universe  (see, e.g., Shapiro  \& Teukolsky 1983). 

Let us denote central density of the $M_{\rm min}$ configuration by 
$\rho_{\rm c,min}$. Configurations with $\rho_{\rm c}\simeq 
\rho_{\rm c,min}$ are very loosely bound by gravitational forces 
(see below). Their  radii are $\sim$ hundreds of kilometers, and a small
difference  in their 
mass is accompanied by a large difference in radii.  
For example, on a stable branch on Fig.\ref{MRmin}, 
a decrease in mass by  $0.001~{\rm M}_\odot$ (i.e., 
by one percent)  implies an increase in radius by $\sim$  
hundred kilometers (i.e., by more than a factor of two !). 
Simultaneously, as seen in 
Fig.\ref{Mrhoc},  for $\rho_{\rm c}<\rho_{\rm c,min}$ (unstable 
branch)  the increase of mass 
with decreasing central density becomes extremely steep.

This behavior is related to  the specific 
structure of neutron stars near $M_{\rm min}$, and to  the 
properties of the EOS of neutron star matter at subnuclear density. 
Neutron stars close to $M_{\rm min}$ are loosely bound,  huge 
spheres of the solid crust, of the radius of hundreds of 
kilometers, 
containing a tiny liquid core, which at $M_{\rm min}$ 
constitutes 2\% of the stellar mass for the SLy EOS, and 
3\% of mass in the case of the FPS model. 

In the context of the stability of equilibrium configurations, 
a particular role is played by the adiabatic index of the 
neutron star interior, 
$\Gamma=(n_{\rm b}/P)({\rm d}P/{\rm d}n_{\rm b})$. As we see in 
Fig.\ref{Mrhoc} the  most dramatic steepening of the $M(\rho_{\rm c})$ 
curve takes place at $\rho_{\rm c}\simeq 1.6~10^{14}~{\rm g~cm^{-3}}$ 
for the SLy EOS, to be compared with $\rho_{\rm c}\simeq 2.2~10^{14}~
{\rm g~cm^{-3}}$ for the FPS EOS. At such central density, neutron stars 
have a tiny, stiff,  liquid core with  $\Gamma_{\rm core}\simeq 2$. 
The average value of the adiabatic index within the crust is much lower,  
$\Gamma_{\rm crust}\simeq 1.3$, below the Newtonian threshold 
for instability with respect to small radial perturbation. 
Gravitational pull barely binds the 
equilibrium configurations, which are unstable with respect to small 
radial perturbations. 

At $M_{\rm min}$, the radius of the liquid core is 3.8 km for the 
SLy EOS, and 4.2 km for the FPS model. As the SLy EOS is stiffer 
for $10^{13}~{\rm g~cm^{-3}}<\rho<\rho_0$ than the FPS one, 
the SLy neutron 
stars at $M_{\rm min}$ are  less compact, and contain smaller and less 
massive liquid core. Parameters of neutron star models with 
minimum mass are collected in Table 1.  

For $\rho_{\rm c}<\rho_{\rm c,min}$, both the mass of the liquid core 
and its radius decrease very slowly with decreasing $\rho_{\rm c}$, while 
the total mass  and total radius of the star increases extremely steeply.  
\section{Binding energy of low-mass  neutron stars}
\begin{figure}
\resizebox{\hsize}{!}{\includegraphics{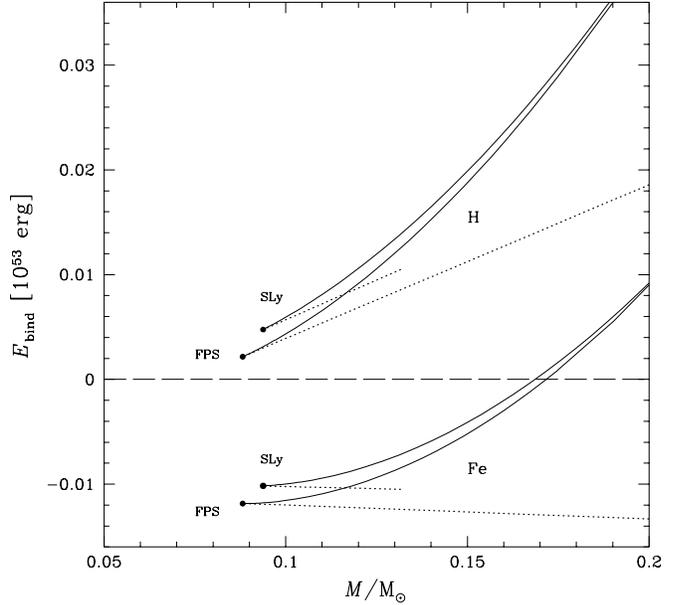}}
\caption{
Gravitational mass versus central density, and binding energy  
 (relative to dispersed $^{56}{\rm Fe}$ - labeled by Fe,  and relative to 
hydrogen gas - labeled by H) versus gravitational mass, 
for cold, static neutron stars. 
Calculations performed for the SLy 
and the FPS EOS. Solid line - stable configurations, dotted lines - 
configurations unstable with respect to small radial perturbations. 
Minimum mass configuration is indicated with a filled circle.  
  }
 \label{MEb}
\end{figure}
%
Binding energy of a neutron star, $E_{\rm bind}$, is defined as the 
mass defect with respect to dispersed configuration of the same 
number of baryons,   
multiplied by $c^2$. Dispersed configuration is characterized by 
negligible pressure and gravitational interactions. 
Equivalently, one may define $E_{\rm bind}$ as a net work  
needed to transform a neutron star into a dispersed configuration. 

One may contemplate several definitions  of dispersed configuration. 
Let us first consider the net work  needed to transform neutron star, 
consisting of $A$ baryons, into a dispersed configuration, 
under the condition of 
 keeping matter always in the 
ground state. In this way, dispersed 
configuration will be a pressureless 
cloud of the dust of $^{56}{\rm Fe}$, with mass per nucleon 
$m_{\rm Fe}\equiv {\rm mass~of~^{56}{\rm Fe}~atom/56}=
1.6587~10^{-24}$~g. Binding energy with respect to such 
a dispersed 
configuration is 
%
\begin{equation}
E^{({\rm Fe})}_{\rm bind}=
\left(Am_{\rm Fe}-M\right) c^2~.
\label{Ebind.Fe}
\end{equation}
%
Another dispersed configuration is that of a pressureless 
hydrogen cloud. Binding energy in that case is
%
\begin{equation}
E^{({\rm H})}_{\rm bind}=
\left(Am_{\rm H}-M\right) c^2~, 
\label{Ebind.H}
\end{equation}
%
where the mass of the hydrogen atom $m_{\rm H}=1.6735~10^{-24}~$g. 

Binding energy of low-mass, cold, static 
 neutron stars, versus stellar mass, in the 
vicinity of $M_{\rm min}$, is plotted in Fig. \ref{MEb}. As we see, 
neutron stars with $M<0.17~{\rm M}_\odot$ are unbound with respect to 
$^{56}{\rm Fe}$. However, neutron stars are always bound with respect 
to the hydrogen cloud configuration. At the same mass (or baryon number), 
unstable configuration (dotted line in Fig.\ref{MEb}) is less bound than 
the stable one. 

The case of $E_{\rm bind}<0$ deserves a comment. Negative binding 
energy with respect to dispersed $^{56}{\rm Fe}$ indicates, that
neutron stars with $M<0.17~{\rm M}_\odot$ are actually 
{\it metastable} with respect to transformation into a 
cloud of $^{56}{\rm Fe}$ dust. Their decompression (while 
locally keeping the matter in the ground state) would eventually 
{\it lower} the total energy, after overcoming an initial energy 
barrier resulting from gravitational binding of the outer layers. 
While being metastable with respect to 
{\it large amplitude} decompression, configurations with $\rho_{\rm c}
>\rho_{\rm c,min}$ are, however, {\it stable} with respect 
to {\it small amplitude} radial perturbations; this is to be 
contrasted with an {\it instability} of configurations with 
$\rho_{\rm c}<\rho_{\rm c,min}$. One expects, that equilibrium 
configurations with $\rho_{\rm c}<\rho_{\rm c,min}$, after being 
perturbed, will eventually expand to huge white-dwarf like configurations 
with radii exceeding $10^4$ km.  
\section{Effect of elastic strains}
Configurations of hydrostatic equilibrium were calculated using 
the Oppenheimer-Volkoff equation (OV)  of general relativity 
(see, e.g., Shapiro \& Teukolsky 1983). The OV equation is    
derived under the assumption that matter is described by the ideal
fluid stress tensor. However, at $M\sim 0.1~{\rm M}_\odot$ a neutron star 
is nearly all solid, except for a small liquid core containing only 
$2-3\%$ of stellar mass. In contrast to liquid, a solid can 
sustain a {\it shear strain}, which contributes to the stress tensor. 
In the isotropic approximation,  the elastic 
shear-strain  term in the stress tensor of the  
Coulomb crystal of the solid crust is determined by the 
shear modulus $\mu$.  
The effect of elastic strain 
on the stellar structure 
can be roughly estimated using the model of Baym \& Pines 
(1971). In the presence of elastic shear strain, 
 static neutron star configuration 
can be non-spherical. Assuming for simplicity that the deviation from 
sphericity is quadrupolar, let us estimate possible   
ellipticity  $\delta R/R$, where $\delta R$ is the 
absolute value of the difference between the polar and 
the equatorial radii.  
For a Coulomb crystal the maximum (i.e., breaking)
strain  does not exceed  $\sim  10^{-2}$ (Ruderman 1992).   
The Baym-Pines (1971) model predicts that  
$\delta R/R < 0.01  b$ where the ``rigidity parameter'' 
$b\equiv \beta \int_{V_{\rm crust}} \mu {\rm d}V/\vert E_{\rm grav}\vert$. 
Here, $E_{\rm grav}$ is the gravitational energy of a neutron star 
 and $\beta$ is a numerical factor $\sim 1$. For a $1.4~{\rm M}_\odot$ 
neutron star, one typically has $b\sim 10^{-5}$, but 
for $M\simeq 0.1{\rm M}_\odot$ one obtains $b\sim 10^{-2}$ 
(Carlini \& Treves 1989). 
Therefore, we expect that the  effect of elastic shear 
strain on the stellar radius 
at $M\sim 0.1~{\rm M}_\odot$ is $\delta R/R< 10^{-3}$. 
\section{Effects of rotation}
\begin{figure}
\resizebox{\hsize}{!}{\includegraphics[angle=-90]{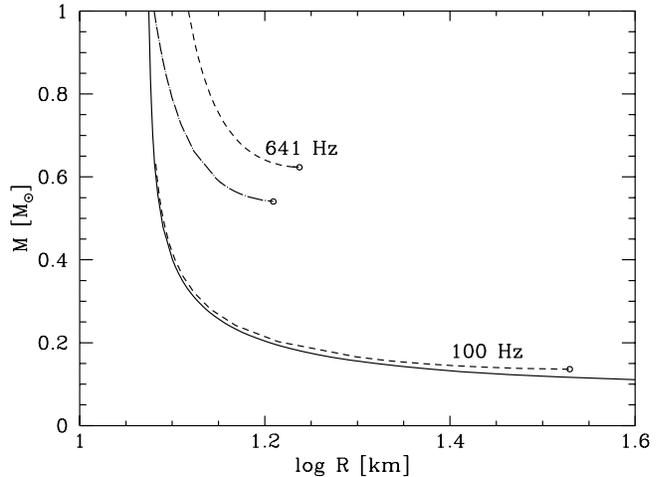}}
\caption{
Gravitational mass, $M$,   versus equatorial 
radius for low-mass neutron stars 
and effects of uniform rotation.  For non-rotating neutron stars 
 equatorial and polar radii are equal, $R_{\rm eq}=R$. 
Solid line corresponds to non-rotating neutron-star models 
based on the SLy EOS. Short-dash lines correspond to uniformly 
rotating neutron-star models based on the SLy EOS, the upper 
one for rotation frequency of the most rapid observed 
millisecond pulsar, and the lower one to rotation period of 
10 ms. Long-dash - dot line describes $M-R_{\rm eq}$ relation
for neutron stars based on the FPS EOS, rotating at the maximum 
observed pulsar frequency. Curves for rotating neutron star models 
terminate at the low-mass end 
 at the Keplerian (mass shedding) configuration, denoted by 
an open circle.  
  }
 \label{MRrot}
\end{figure}
The shortest observed pulsar period is $P_{\rm min}^{\rm obs}=
1.558~$ms, which corresponds to rotation frequency 
$\nu_{\rm max}^{\rm obs}=641~$Hz. For massive neutron stars 
of $M\ga 1.3~{\rm M}_\odot$, rotation at such frequency 
has a rather small offect on neutron star structure, and 
can  be described using {\it slow rotation approximation} 
(Hartle 1967). Leading effects of rotation on neutron star 
structure are then quadratic in $\nu$, and are small as long 
as gravitational pull, acting on a unit mass element 
 at the equator $\sim GM/R^2$ is significantly 
larger than the centrifugal force $\sim 4\pi^2 R \nu^2$. 
However, with decreasing mass of a neutron star rotating at 
fixed $\nu$, the rotational effects 
become more and more important. At the same  $M$, the equatorial radius of 
a rotationally flattened neutron star becomes significantly larger 
than that of a static configuration. 
Importance of rotation increases with increasing equatorial radius 
and decreasing mass, and become decisive for $\nu^2\sim 
GM/(4\pi^2 R^3)$. 

We performed exact calculations of stationary configurations of 
cold neutron stars, rotating uniformly at  100 Hz and 641 Hz. 
Numerical calculations were carried out  using exact 2-D equations 
for uniformly rotating configurations in general relativity. 
Equations of hydrostatic equilibrium were solved using 
numerical code, based on the pseudospectral method, and developed 
at DARC Meudon (Bonazzola et al. 1998, Gourgoulhon et al. 1999). 
Low-mass  constant-$\nu$ sequences in the $M-R$ plane are 
displayed in Fig. 5.

Effects of rotation are larger for the stiffer SLy EOS, which is 
most natural. At 641 Hz and $M<0.8~{\rm M}_\odot$, effects of 
rotation are very large. The lowest-mass configuration corresponds to 
the mass-shedding limit. We get 
$M_{\rm min}^{\rm SLy}(641~{\rm Hz})=0.61~{\rm M}_\odot$. 
For the FPS EOS, the effect is smaller, because it is softer than 
the SLy one, and therefore low-mass neutron stars  are more compact. 
We obtain 
$M_{\rm min}^{\rm FPS}(641~{\rm Hz})=0.54~{\rm M}_\odot$. 
Even at $\nu=100$ Hz, rotation has a sizable effect, 
with minimum mass, reached at the mass-shedding limit, 
$M_{\rm min}^{\rm SLy}(100~{\rm Hz})=0.13~{\rm M}_\odot$, 
larger,  by nearly 40\%,  than that for static neutron stars.

For the  SLy minimum mass configuration  rotating at
$\nu=641$~Hz the equatorial and polar radius of the liquid core are
11 km and 9 km,  respectively. 
However, the  crust is much more deformed  than the whole star,
 having
equatorial thickness of 6 km,  four times larger than the polar one. 
This results in quite 
a large oblateness, $R_{\rm pole}/R_{\rm eq}=0.6$. 
The mass  is concentrated in a liquid core, which 
contains 93\% of neutron star mass.

Minimum mass configuration, calculated for the SLy EOS at 
 $\nu=100$~Hz, has  the (nearly spherical) liquid core
 constituting  half of the stellar mass. However, the 
equatorial radius of the liquid core 
is  only about 
$6$~km,  to be compared with the 28 km  equatorial thickness 
of the crust. Consequently, the oblateness of the minimum mass 
configuration is large, $R_{\rm pole}/R_{\rm eq}\simeq 0.7$.

Our calculations show, that 
the rotating configurations  close to minimum mass cannot 
be approximated by
a  homogeneous rotating ellipsoid, like the model of 
Colpi et al. (1991).  The two  quantities relevant for   
the Keplerian (mass shedding) limit, mass and equatorial radius, 
are determined by different parts of a low-mass neutron star. 
Namely, the mass is supplied  by the liquid core, while the equatorial 
radius is defined by  the  outer edge of the 
extended solid crust. Therefore, 
the approximation 
in which a neutron star is represented by a homogeneous 
(constant density - incompressible) spheroid (Colpi et al. 1991) is 
inadequate. 

 It should be mentioned, that even at the mass-shedding limit 
at 641 Hz, rotating configurations are not susceptible to 
 the viscosity-driven tri-axial instability which would imply 
emission of gravitational radiation.  
 This secular instability sets in at kinetic to gravitational 
potential energy ratio $T/W=0.14$. 
We get $T/W=0.061$ 
for the SLy EOS, and even less, $T/W=0.056$, for the softer 
FPS EOS. Our conclusion  is opposite to that obtained by 
Colpi et al. (1991), which 
 results from the inadequacy of the model used by these authors.

 One may ask, whether  the values $T/W=0.05-0.06$ are  above 
the threshold for the  gravitational-radiation driven   
Chandrasekhar-Friedman-Schutz (CFS) instability. The threshold 
values  $T/W\sim 0.04-0.05$ for the $m=4,5$ CFS instabilities, 
calculated for realistic EOS by 
Morsink et al. (1999),   correspond to  $M\ga
0.8~{\rm M}_\odot$.   
Extrapolation of these results to 
 strongly deformed  
configurations with $M\sim 0.6~{\rm M}_\odot$  rotating at 641 Hz (Fig. 5) 
would be risky. Therefore, the problem of the CFS instabilities 
in rotating low-mass neutron stars has to be clarified in the 
future studies.

One has to point out basic difference between low-mass  stationary 
uniformly 
rotating configurations and the static ones. Let us denote central 
density of the minimum mass configuration by 
$\rho_{{\rm c}, M_{\rm min}}$. 
In the static case, equilibrium 
configurations with 
$\rho_{\rm c}<\rho_{{\rm c}, M_{\rm min}}$ are secularly unstable
 with respect  to radial perturbations. However, for 
$\nu \ga 100$ Hz, stationary configurations with 
$\rho_{\rm c}<\rho_{{\rm c}, M_{\rm min}}$ 
just do not exist. 
\section{Summary  and conclusion}
In the present paper we determined the properties of the 
minimum mass configuration of neutron stars, built of 
cold catalyzed matter. We used two 
recent EOS of cold catalyzed matter, which describe in a 
unified way both the crust and the liquid core, as well as their 
interface. For static neutron stars, we get a minimum mass of $0.09~
{\rm  M}_\odot$,  coinciding with  
the value of $M_{\rm min}$ obtained three decades ago in the 
BPS paper. 
 The value of $M_{\rm min}$ for 
neutron stars  built of cold catalyzed matter is quite independent 
of the specific EOS used.  However, the structure of the minumum 
mass configuration is sensitive to the EOS at $\rho\la \rho_0$. 
For the  FPS and SLy EOS we get  $R(M_{\rm min})\simeq 220$~km 
and 270 km, 
 respectively. Our minimum mass configurations contain a tiny 
liquid core with $2-3\%$ of the stellar mass. The structure 
of the $M_{\rm min}$ configurations for the EOS considered in the 
present paper are substantially different from that obtained in 
the classical BPS paper. 
While BPS got a similar value of $M_{\rm min}$, the radius of their 
minimum mass configuration was only 160 km, and  it was 
completely solid. These differences can be easily understood in 
terms  of the differences in the EOS at subnuclear density. Namely, 
for $\rho\la \rho_0$, the BPS EOS is softer, and  its crust-core 
transition takes place at significantly higher density than 
for the SLy and FPS ones. 

Within mass interval  $M_{\rm min}=0.09<M<0.17~{\rm M}_\odot$, 
neutron stars are unbound with respect to dispersed $^{56}{\rm Fe}$. 
In the similar mass range, neutron stars are less bound 
than  carbon-oxygen white dwarfs of the same mass. 
However, all neutron stars are bound with respect to dispersed 
hydrogen.

Effects of rotation have been shown to be important for $M_{\rm min}$, 
and for the low-mass neutron stars in general. For neutron stars 
rotating rigidly at $P=1.56~$ms, minimum mass is determined 
by the mass-shedding limit: it is $0.5~{\rm M}_\odot$ 
and $0.6~{\rm M}_\odot$,
for the FPS and SLy EOS, respectively. The minimum mass 
configuration at $P=1.56~$ms is quite  oblate, due 
to strong deformation 
of the crust by the centrifugal force.

\begin{acknowledgements}
We are grateful to A. Potekhin for his help in preparation 
of Figs.1,3,4. We are also very grateful to the referee, 
S. Blinnikov, for his remarks and comments which 
helped to improve the present paper.  
This research
was partially supported by the KBN grant No. 5P03D.020.20.
\end{acknowledgements}


\begin{thebibliography}{} 

\bibitem[1971a]{BBP}
Baym G., Bethe H.A., Pethick C., 1971a, Nucl. Phys. A 175, 225

\bibitem[1971]{BPS}
Baym G., Pethick C., Sutherland P., 1971b, ApJ 170, 299 ({\bf BPS})

\bibitem[1971]{BaymPines}
Baym G., Pines D., 1971, Ann. Phys. 66, 816

\bibitem[1984]{Blinnikov1984}
Blinnikov S.I., Novikov I.D., Perevodchikova T.V., 
Polnarev A.G., 1984, 
Sov. Astron. Letters, 10, 177


\bibitem[1998]{Bona1998}
Bonazzola S., Gourgoulhon E., Marck. J.-A., 1998, 
Phys. Rev. D 58, 104020

\bibitem[1989]{Carlini89}
Carlini A., Treves A., 1989, 
A\&A 215, 283 

\bibitem[1997]{Chabanat1997}
Chabanat E., Bonche P., Haensel P., Meyer J., Schaeffer R., 
1997, 
Nucl. Phys. A 627, 710

\bibitem[1998]{Chabanat1998}
Chabanat E., Bonche P., Haensel P., Meyer J., Schaeffer R., 
1998, 
Nucl. Phys. A 635, 231 

\bibitem[1959]{Cameron1959}
Cameron A.G.W., 1959, 
ApJ 130, 884 

\bibitem[1989]{Colpi1989}
Colpi M., Shapiro S.L., Teukolsky S.A., 1989, 
ApJ 339, 318

\bibitem[1991]{Colpi1991}
Colpi M., Shapiro S.L., Teukolsky S.A., 1991, 
ApJ 369, 422 

\bibitem[1970]{CohenCam1970}
Cohen J.M., Cameron A.G.W., 1971,
ASS 10, 227 

\bibitem[2000]{DH2000}
Douchin F., Haensel P., 2000, 
Phys. Lett. B 485, 107

\bibitem[2001]{DH2001}
Douchin F., Haensel P., 2001, 
A\&A 380, 151 

\bibitem[2000]{DHM2000}
Douchin F., Haensel P., Meyer J., 2000, 
Nucl. Phys. A  665, 419

\bibitem[1999]{Gourgoulhon1999}
Gourgoulhon E., Haensel P., Livine R., Paluch E., 
Bonazzola S., Marck J.-A., 1999,
A\&A 349, 851 

\bibitem[1998]{Goussard1998}
Goussard J.-O., Haensel P., Zdunik J.L., 1998, 
A\&A 330, 1005

\bibitem[2001]{Haensel2001}
Haensel P., 2001,  
Neutron star crusts, in: Physics of Neutron Star Interiors, 
D. Blaschke, N.K. Glendenning, A. Sedrakian (Eds.), 
Springer - Lecture Notes in Physics, Berlin - New York, 
p. 127 

\bibitem[1965]{HTWW65}
Harrison B.K., Thorne K.S., Wakano M., Wheeler J.A., 1965, 
Gravitation Theory and Gravitational Collapse, University 
of Chicago Press, Chicago

\bibitem[1967]{Hartle1967}
Hartle J.R., 1967,
ApJ 150, 1005


\bibitem[1938]{Landau1938}
Landau, L.D., 1938, 
Nature 141, 333

\bibitem[1969]{Langer1969}
Langer W.D., Rosen L.C., Cohen J.M., Cameron A.G.W., 1969,
ASS 5, 259 


\bibitem[1961]{LevSimm1961}
Levinger J.S., Simmons L.M., 1961, 
Phys. Rev. 124, 916 

\bibitem[1993]{Lorenz93}
Lorenz C.P., Ravenhall D.G., Pethick C.J., 1993,  
Phys. Rev. Lett. 70, 379

\bibitem[1999]{Morsink1999}
Morsink S.M., Stergioulas N., Blattnig S.R., 1999, 
ApJ 510, 854

\bibitem[1938]{OppSerber1938}
Oppenheimer J.R., Serber R., 1938, 
Phys. Rev. 54, 540

\bibitem[1992]{Ruderman1992}
Ruderman M., 1992, Neutron star crust breaking and magnetic field 
evolution, in: The Structure and Evolution of Neutron Stars, 
D. Pines, R. Tamagaki, S. Tsuruta (Eds.), Addison-Wesley Publishing 
Company

\bibitem[1983]{ST1983}
Shapiro S.L., Teukolsky S.A., 1983, 
Black Holes, White Dwarfs and Neutron Stars, Wiley, 
New York


\bibitem[1998]{Sterg1998}
Stergioulas N., Friedman J.L., 1998, 
ApJ 492, 301

\bibitem[1999]{Strobel1999}
Strobel K., Schaab C., Weigel M.K.,  1999, 
A\&A 350, 497 

\bibitem[1998]{Sumi1998}
Sumiyoshi K., Yamada S., Suzuki H., Hillebrandt W., 1998, 
A\&A 334, 159

\bibitem[1966]{Tsuruta1966}
Tsuruta S., Cameron A.G.W.1966, 
Canadian J. of Phys. 44, 1895 
\end{thebibliography}
\end{document}